\batchmode
\documentstyle[aaspp4,12pt]{article}
\begin{document}
\newcommand{\simlt}{\lesssim}
\newcommand{\simgt}{\gtrsim}
\newcommand{\msol}{M_{\sun}}
\newcommand{\rhon}{\rho_{\rm n}}
\newcommand{\sign}{\sigma_{\rm n}}
\newcommand{\nn}{n_{\rm n}}
\newcommand{\cc}{\rm{cm}^{-3}}
\newcommand{\muci}{\mu_{\rm{c0}}}
\newcommand{\lref}{l_{\rm{ref}}}
\newcommand{\nnci}{n_{\rm{n,c0}}}
\newcommand{\xxi}{x_{\rm{i}}}
\newcommand{\xxici}{x_{\rm{i,c0}}}
\newcommand{\Beqci}{B_{z,\rm{c0}}}
\newcommand{\Beq}{B_{z}}
\newcommand{\nni}{n_{\rm{i}}}
\newcommand{\Myr}{{\rm{Myr}}}
\newcommand{\nncrit}{n_{\rm{n,c,crit}}}
\newcommand{\tcrit}{t_{\rm{crit}}}
\newcommand{\tevol}{t_{\rm{evol}}}
\newcommand{\tlife}{\tau_{\rm{life}}}
\newcommand{\tdetect}{\tau_{\rm{det}}}
\newcommand{\tammonia}{t_{\rm{NH_{3}}}}
\newcommand{\lamTcrit}{\lambda_{\rm{T,c,crit}}}
\newcommand{\nmean}{\langle \nn \rangle}
\newcommand{\vd}{v_{\rm D}}
\newcommand{\vnr}{v_{{\rm n},r}}
\newcommand{\vir}{v_{{\rm i},r}}
\newcommand{\Bref}{B_{\rm ref}}
\newcommand{\muG}{\mu{\rm G}}
\newcommand{\vn}{v_{\rm n}}
\newcommand{\vi}{v_{\rm i}}
\newcommand{\ul}{\underline{\hspace{40pt}}}
\title{CONSISTENCY OF AMBIPOLAR DIFFUSION MODELS WITH INFALL IN THE L1544
PROTOSTELLAR CORE}
\author{Glenn E. Ciolek\altaffilmark{1} and Shantanu Basu\altaffilmark{2}\altaffilmark{, 3}}
\altaffiltext{1}{New York Center for Studies on the Origin of Life (NSCORT),
and Department of Physics, Applied Physics, and Astronomy,
Rensselaer Polytechnic Institute, 110 8th Street, Troy, NY 12180;
cioleg@rpi.edu.}
\altaffiltext{2}{Canadian Institute for Theoretical Astrophysics,
University of Toronto, 60 St. George Street, Toronto, Ontario M5S 3H8,
Canada.}
\altaffiltext{3}{Current address: Department of Physics and Astronomy,
University of Western Ontario, London, Ontario N6A 3K7, Canada; 
basu@astro.uwo.ca.}
\vspace{3ex}
\begin{center}
Accepted for publication in {\it The Astrophysical Journal},
Vol. 529 \#2, 1 February 2000
\end{center}
\begin{abstract}
Recent high-resolution studies of the L1544 protostellar core 
by Tafalla et al. and Williams et al. reveal the structure and the 
kinematics of the gas. The observations of this prestellar
core provide a natural test for theoretical models of core formation
and evolution. Based on their results, the above authors claim a 
discrepancy with the implied infall motions from ambipolar diffusion models.
In this paper, we reexamine the earlier ambipolar diffusion models,
and conclude that the L1544 core {\it can} be understood to be a
magnetically supercritical core undergoing magnetically diluted
collapse.  We also present a new ambipolar diffusion model specifically
designed to simulate the formation and evolution of the L1544 core.
This model, which uses reasonable input parameters, yields 
mass and radial density distributions, as well as neutral and
ion infall speed profiles, that are in very good agreement
with physical values deduced by observations.
The lifetime of the 
core is also in good agreement with mean prestellar core lifetimes 
estimated from statistics of an ensemble of cores. 
The observational input can act to constrain other currently
unobserved quantities such as the degree of ionization, and the
background magnetic field strength and orientation near the L1544 core.

\end{abstract}
\section{Introduction}
That only a very small fraction of the mass in molecular clouds is
ultimately converted into stars is well established by
observations (e.g., Zuckerman \& Palmer 1974; Carpenter et al. 1993;
Fuller 1994). Mouschovias (1976, 1977, 1978)
proposed that the observed inefficiency in star formation
is naturally explained by molecular clouds being supported against
gravitational collapse by interstellar magnetic fields. Collapse in the
interior flux tubes of a cloud is initiated by ambipolar diffusion, the
gravitationally-induced drift of neutral gas with respect to plasma and
magnetic field.
Extensive discussion of various aspects of star formation
in magnetic clouds, including the role of ambipolar diffusion, can be found
in the reviews by Mouschovias (1987), Shu, Adams, \& Lizano (1987), McKee et al.
(1993), and Mouschovias \& Ciolek (1999). 

Early axisymmetric calculations considered the initial quasistatic
evolution driven by ambipolar diffusion (Mouschovias 1979; Nakano 1979;
Lizano \& Shu 1989).  Fiedler \& Mouschovias (1992, 1993) presented
detailed numerical simulations showing the formation and subsequent
collapse of supercritical cores in two-dimensional, axisymmetric, isothermal,
self-gravitating, magnetic model molecular clouds. The flattening of
the cloud along the magnetic field lines in these models was
used in companion studies by Ciolek \& Mouschovias (1993, 1994;
hereafter CM93 and CM94, respectively), who modeled the formation of
cores by ambipolar diffusion in disk-like magnetic clouds, including
the effects of charged and neutral dust grains in their calculation.
Disk-like model clouds were also used by Basu \& Mouschovias
(1994, 1995a, b; hereafter BM94, BM95a, BM95b), who studied the roles of
rotation and magnetic braking in the evolution of cores. The effect
of an external ultraviolet radiation field, which ionizes the outer
regions of a cloud, was incorporated into these models by Ciolek
\& Mouschovias (1995; hereafter CM95). Qualitatively, the results of these
calculations were quite alike: supercritical cores formed due to
ambipolar diffusion and went on to collapse dynamically while embedded
in magnetically supported (subcritical) clouds.
These studies, as well as other magnetic models without ambipolar diffusion
(e.g., Tomisaka 1996; Li \& Shu 1996; Nakamura et al. 1999), show that
collapsing supercritical regions should have an oblate shape.

Past comparison of observed cores with ambipolar diffusion models has
yielded positive results. A model for the Barnard 1 cloud was presented 
by Crutcher et al. (1994); core properties predicted by their model 
(such as the mass, mean density, and magnetic field strength) were found to be
in excellent agreement with observed values (in some instances, differing by less
than 10\%). In addition, sub-mm continuum observations of prestellar cores
(Ward-Thompson et al. 1994; Andr\'{e}, Ward-Thompson, \& Motte 1996), yield
density profiles that are in reasonable agreement with those developed in the ambipolar
diffusion models. Benson, Caselli, \& Myers (1998) measured the
velocity difference between $\rm{NH_{3}^+}$ and neutral molecules
($\rm{C_{3}H_{2}}$ and CCS) in sixty dense cores; they noted that their
upper limit on the ion-neutral drift speed
($\simeq 0.03~\rm{km}~{\rm s}^{-1}$) is consistent with our 
published models of core formation and contraction due to ambipolar
diffusion. 

The L1544 cloud, located in Taurus, seems to be a particularly well-suited
candidate for comparison with theoretical models. 
\footnote{This is especially true since surveys by Moneti et al.
(1984), Goldsmith  \& Sernyak (1984), and Heyer et al. (1987) have
provided ample indirect evidence to suggest that the large-scale
evolution of the Taurus dark clouds is magnetically controlled.}
Ward-Thompson et al. (1994) suggested that the L1544 core could be a
transitional object between a starless core and a core with an IRAS
source. As such, the evolutionary state of L1544 could correspond to a
recently formed supercritical core which is entering the stage of rapid
infall, according to the magnetic models described above. Recent
observations by Tafalla et al. (1998, hereafter referred to as T98)
indicate that L1544 does show infall on scales of 0.1 pc. More recently,
Williams et al. (1999, hereafter referred to as W99)
have presented a high-resolution interferometric study of the L1544
core, which allows them to map the kinematics of the ionized species
$\rm{N_{2}H^{+}}$ down to radii $\sim$ 0.02 pc. In both papers,
significant infall motions are detected and used to
call into question the applicability of the ambipolar diffusion models
presented by, e.g., CM95 and BM94. A further point in favor
of relatively rapid infall motions is the relative statistics
of cores with and without embedded protostars, implying a starless
core lifetime of $\sim$ 1 Myr (e.g., Lee \& Myers 1999).

In this paper we investigate whether ambipolar diffusion models are
relevant in explaining the formation and evolution of the L1544 core.
In \S~2 we reexamine the results of the earlier ambipolar diffusion models 
in light of the new observational surveys.
We find that, in reality, many of the results of the
earlier models are consistent with the infall deduced from observations,
if the infall region is taken to represent a portion of the supercritical core.
In \S~3 we present a new ambipolar diffusion model specifically designed 
to simulate the
evolution of the L1544 core. This particular model uses as input data 
quantities well within the observationally allowed region of physical 
parameters.
Our results are summarized in \S~4.
\vspace{-4ex}
\section{Comparison of Earlier Ambipolar Diffusion Models With Recent Observations}
\vspace{-2ex}
A key result of axisymmetric numerical simulations of magnetically
regulated star formation (Fiedler \& Mouschovias 1993; CM94, CM95;
BM94, BM95a, BM95b) is 
the formation of an inner region with supercritical mass-to-flux ratio.
A supercritical region can only form when the central column density 
(indirectly, the density as well) exceeds a critical value effectively 
determined by the background magnetic field strength.
A supercritical region with size proportional to the
thermal critical length scale (see \S~3) at the critical column density
begins rapid collapse and effectively separates
from its subcritical surroundings. Scaling laws for the sizes and masses of
such cores are given by BM95b. 
The cores are characterized by power-law profiles in density, magnetic
field components, and angular velocity. The innermost regions have
density profiles approaching $r^{-2}$, but the outer regions (where most of
the mass is contained) have progressively shallower slope, so that if we write
$\rhon \propto r^{s}$, the mass-weighted mean $\bar{s}$ falls 
between $-1.5$ and $-1.7$ for the various models 
(CM94, CM95; BM94, BM95a, BM95b). The shallowness 
of the outer profile is due to increasing magnetic
support (alternatively, decreasing mass-to-flux ratio) in the outer core.
The cores are flattened due to the remaining significant magnetic
support, with typical axial ratios in the range $0.25 - 0.33$. 
An individual observed core is best compared with theory while keeping the
whole ensemble of models in mind (e.g., see the parameter study done by 
BM95a, BM95b), rather than comparing with a single standard model.
This is due to the considerable observational
uncertainty in input parameters such as the background magnetic field strength
and ionization fraction, as well as real variations allowed by nature.
Below, we interpret the L1544 observations in terms
of an overall understanding gained from the parameter studies. 

A point of general agreement of the models with observations is the
conclusion of T98 that the overall density profile in L1544 is
shallower than $r^{-2}$. They find an approximate fit with
an $r^{-1.5}$ profile over a scale of 0.1 pc. Several simplifying assumptions
are used to obtain this estimate (see T98), and error bounds are not explicitly
estimated, although $r^{-1}$ and $r^{-2}$ profiles are ruled out.
We point out that an approximate $r^{-1.5}$ density profile in a prestellar core
is inconsistent with a scale-free isothermal ball of gas, but {\it is}
in reasonable agreement with the magnetic models discussed above. 
An apparent disagreement with the magnetic models as stated by T98
and W99 is the evidence for extensive inward motions, of order 0.1 
km s$^{-1}$, from scales of 0.1 pc (T98) down to 0.02 pc (W99).
The claim that these motions exceed that of the magnetic models over
similar ranges in length and density is based on a comparison with the
standard models of the above studies.

We first compare the W99 results with the existing models, since infall
motions on the small scales $\sim$ 0.02 pc are more easily produced in
these models. The essence of W99's claim is that at radius $\approx$
0.02 pc, where the inferred density is $\approx 4 \times 10^5$ cm$^{-3}$,
the inferred infall motions for the ionic species $\rm{N_{2}H^{+}}$, 
$\approx 0.08 \rm{km}~{\rm s}^{-1}$, greatly exceed those of ambipolar
diffusion models. We refer the reader to W99 for an exposition of the
significant uncertainties in 
obtaining these numbers. For the remainder of this paper, we take these
numbers at face value. When we examine the standard model 2 of BM94
and the standard $\rm{B_{UV}}$ model of CM95 (scaling the results to
the estimated kinetic temperature 12 K), we find that the infall is
indeed slower than these numbers when the {\it central} density is
close to $3 \times 10^5$ cm$^{-3}$, i.e., BM94's model 2 has ionic infall 
velocity $\vir = -0.012$ km s$^{-1}$ and CM95 has $\vir = -0.010$ km s$^{-1}$.
However, the observations, which have spatial resolution $\approx$
0.01 pc, cannot determine the central density. Therefore, it is prudent
to compare the evolutionary models at all possible times when the 
density at radius $\approx 0.02$ pc is in approximate agreement with
the estimated density $\approx 4 \times 10^5$ cm$^{-3}$.
It is striking that the development of a power-law density profile in 
the magnetic models yields an asymptotic value of the density at this
radius that is very close to the observed one. At later times, when
supercritical collapse is well developed, the density at $r = 0.02$ pc
is $9 \times 10^5$ cm$^{-3}$  and $6 \times 10^5$ cm$^{-3}$ in the BM94
and CM95 models, respectively
\footnote{Another interesting property of these distributions (and
power-laws in general) is that the mean value within any radius is
of the same order as the value at that radius; typically just a few 
times greater.}. 
At the same radius, the asymptotic values of $\vir$
are $-0.11$ km s$^{-1}$
and $-0.07$ km s$^{-1}$ in the same two models. Since the likely
inclination angle $\theta$ for the disk is $15 \arcdeg - 30 \arcdeg$
(see \S\ 3), the $\cos \theta$ factor due to the inclination of the
disk can reduce the maximum line of sight velocity by a factor of only
$0.87-0.96$. Hence, we conclude that the infall observations of W99 are
{\em not} in disagreement with the standard magnetic collapse models if
the cores are in a later stage of development, with an unresolved
central density $\gg 10^5$ cm$^{-3}$. Such models are typically
characterized by background magnetic field strengths of
$\Bref \approx 30 \muG$ and ionization fraction $x_i \approx 10^{-7}$
at a neutral density $\approx 10^4$ cm$^{-3}$. Indeed, W99 themselves
comment that their results may be in agreement with the rapid infall
observed in the model of Li (1998) if late time collapse is considered. 
We are in agreement with this
statement since the model of Li (1998), while restricted to an unnatural
spherical geometry in the presence of magnetic fields, and lacking
the quantitatively important magnetic tension force\footnote{Due to its
nature, such a model cannot predict the magnetic field geometry or the
aspect ratio of the core.},
reproduces qualitatively the broad features of infall found in the 
more detailed models described above.

The evidence for a later stage of evolution is strengthened by the 
millimeter dust continuum measurements of Ward-Thompson, Motte, \&
Andr\'{e} (1999), whose high-resolution observations establish a
flattening of the column density profile of L1544 on inner scales of
$\sim 10^3$ AU, with a central column density
$N \approx 10^{23}$ cm$^{-2}$. This central column density places the
evolutionary state of the core between the times $t_3$ and $t_4$ in
most of our models (see \S\ 3), corresponding to a central density in
the range $3 \times 10^6 - 3 \times 10^7 \; \cc$.
 
On larger scales of $\sim 0.1$ pc, T98 detect infall motions with
maximum values $\approx$ 0.1 km s$^{-1}$. These motions {\it are}
somewhat greater than the infall speeds presented in the standard
models mentioned above. At a radius $r = 0.1$ pc, the standard model 2
of BM94 and the $\rm{B_{UV}}$ model of CM95 yield neutral infall speeds
of $-0.06$ km s$^{-1}$ and $-0.04$ km s$^{-1}$, respectively (again,
for $T = 12$ K). The parameter study of BM95a and BM95b reveals two
general means by which this factor $\sim 2$ (or slightly higher due to
disk inclination) discrepancy can be resolved: (1) A lower background
magnetic field strength, yielding a lower critical density for collapse
and consequently a larger supercritical core (see BM95b, models 5 and
6). (2) A lower ionization fraction, yielding relatively more rapid
infall at all times (see BM95a, models 8 and 9). Both magnetic field
strengths and ionization fractions are characterized by appreciable
observational uncertainty. In this paper, we concentrate primarily on
the first possibility; a lower background magnetic field strength.
Model 6 of BM95b can be interpreted as having a $\Bref$ of about $10
\muG$. The central flux tube initially has a critical mass-to-flux
ratio, although the rest of the cloud is initially subcritical. Results
of this model, presented in BM95b (see their Fig. 6) shows an extended
infall zone due to the low
density at which a supercritical core is formed. Using $T=12$ K yields
neutral and ionic infall velocities $\vnr = -0.11$ km s$^{-1}$ and
$\vir = -0.09$ km s$^{-1}$ at $r= 0.1$ pc. Significant infall is
actually present at even larger radii in this model.

Our point in reviewing these models is not that any single one fits the 
observations of L1544 exactly, but that the observed features are broadly
consistent with features found in one or many previously published models
which use reasonable input parameters. Therefore, the  L1544 core can very 
likely be modeled as a supercritical core undergoing magnetically diluted 
collapse. In the following section, we present a new model which can
simultaneously match as many features of the L1544 core as possible.
\vspace{-4ex}
\section{An Ambipolar Diffusion Model For L1544}
\vspace{-2ex}
We now turn to a model designed to more accurately simulate inferred
quantities for the L1544 core. Dimensionless parameters used as initial
input for this model are: a central mass-to-flux ratio (in units of the
critical value for collapse) $\muci=0.80$, and a reference state column
density lengthscale $\lref = 7.5 \pi$. (The definition and meaning of
these and the other free parameters specifying our disk models are
discussed in CM93, CM94, CM95, and BM94, BM95a, b.) Since there have
been no significant azimuthal velocities detected in L1544, we neglect
for now the effect of rotation and magnetic braking. We do account for
the effect of an external ultraviolet radiation field acting on the
cloud; the dimensionless parameters $\zeta_{\alpha_{0},\rm{UV,CR}}$
($\equiv \zeta_{\alpha_{0},\rm{UV}}/\zeta_{\rm{CR}}$, where
$\zeta_{\alpha_{0},\rm{UV}}$ is the UV ionization rate of neutral atomic
species $\alpha_0$ and $\zeta_{\rm{CR}}$ is the cosmic-ray ionization
rate), which fix the UV ionization rates at the cloud boundary, are the
same as those listed for models $\rm{A_{UV}}$ and $\rm{B_{UV}}$ in
Table 1 of CM95, except that they are to be multiplied by a factor
= (5/1.3)= 3.8, reflecting our use of a cosmic-ray ionization rate
$\zeta_{\rm{CR}}=1.3 \times 10^{-17}~{\rm s}^{-1}$ instead of the
$\zeta_{\rm{CR}}=5 \times 10^{-17}~{\rm s}^{-1}$ used by CM95.
As in our previous models, the dimensionless cloud radius $\xi_{R}$ is
taken to be $\gg \lref$. We select a value of $\xi_{R}$ such that
a mass $\simeq 30~\msol$ is contained within the dimensional radius
$r \simeq 0.45~\rm{pc}$, as estimated for the L1544 cloud by T98 (see 
their Fig. 1). The mass contained at radii
$\simgt 0.5$ pc, which is required in our models solely for
computational purposes --- the gravitational field diverges if we
abruptly cut off the mass distribution at this radius (see the
associated discussion on this point in \S~4.1 of Crutcher et al. 1994)
--- has little effect on the dynamics of the core in our model cloud,
and remains essentially fixed in space because of effective magnetic
support at large radii (see Figs.  $1a$-$1c$). Finally, the collisional
effects of grains have been made negligible by assuming relatively
large grains, with radius $a = 10^{-5}$ cm. The larger cross-section
of these grains is more than offset by their low abundance relative
to the ions (CM93, Ciolek \& Mouschovias 1996, 1998). The grains,
however, still play a crucial role in the evolution of our model, in
that recombination of ions and electrons takes place on their surfaces,
which affects the chemical reaction rates and the calculation of the
abundances of charged particles throughout the cloud (CM94,
Ciolek \& Mouschovias 1998).

Physically, the parameters cited above could represent an isothermal
cloud with temperature $T$ = 12 K, initial central density and
magnetic field strength (in the equatorial plane of the cloud)
$\nnci = 4.37 \times 10^3~\cc$ and $\Beqci = 16.5~\muG$,
respectively. Because the cloud is only 20\% subcritical, $\Beqci$ is a
factor 1.35 greater than the reference value $\Bref = 12.2~\muG$, which
is the field strength in the outer portion of the cloud.
The initial central degree of ionization
$\xxici \equiv (\nni/\nn)_{\rm{c0}} =  8.5 \times 10^{-8}$
(where $\nni$ is the ion density).

Qualitatively, the overall ambipolar-diffusion-initiated 
infall that occurs in this model is similar to that which occurred
in our previous numerical simulations, cited in the preceding
sections. The evolution of the density profile in the model cloud is
displayed in Figure $1a$ at eleven different times $t_j$
($j = 0$,1,2,...,10), when the central density
$n_{\rm{n,c}}(t_{j}) = 10^j \nnci$;
these times are, respectively, 0, 2.27, 2.60, 2.66, 2.680, 2.684, 2.685,
2.6856, 2.68574, 2.68577, and 2.68578 $\Myr$.
The central mass-to-flux ratio is equal to the critical value for collapse
at a central density $\nncrit = 9.17 \times 10^3~\cc$ for this model, and is
achieved at a time $\tcrit = 1.30~\Myr$. As in our earlier studies, an
asterisk on a curve indicates the instantaneous position of the critical
magnetic flux tube $R_{\rm{crit}}(t_{j})$ (= radius of the supercritical core). 
For $r < R_{\rm{crit}}$, the mass-to-flux ratio in each flux tube exceeds the
critical value for collapse, $(M/\Phi_{\rm B})_{\rm{crit}} = 1/(4 \pi^2 G)^{1/2}$.
An open circle on a curve locates the central thermal critical lengthscale,
$\lambda_{\rm{T,c}}(t_{j}) = C^2/2 G \sigma_{\rm{n,c}}$, where $C$ is the
isothermal speed of sound [$= 0.19 (T/10~\rm{K})^{1/2}~\rm{km}~{\rm s}^{-1}$],
$G$ is the gravitational constant, and $\sigma_{\rm n,c}$ is the central
column density. The radius of the core that formed in this model is equal to
0.30 pc. The mass contained in the core is $19~\msol$; the mean core density
is $\nmean_{\rm{core}}=8.0 \times 10^3~\cc$, which is lower than the
density for collisional excitation of ammonia ($=1.1 \times 10^4~\cc$).
Note that the core radius is greater than the scale of the observations
made by both W99 (= 0.02 pc) and T98 ($\simeq 0.14$ pc), hence, as
discussed in \S~2, we identify the region studied in these dedicated
studies as existing {\em within} the supercritical core. The logarithmic
derivative of the density $s \equiv \partial \ln \nn/ \partial \ln r$
is presented in Figure $1b$, at the same eleven times $t_j$. Not long
after the formation of the core ($t_{j} > t_{3}$), for scales $r \simgt 0.02~\rm{pc}$,
$s$ is typically $> -1.8$. Figure $1c$ displays the vertical column
density $\sign(r)$ [$= 2 \rhon(r) Z(r)$ in our models, where $\rhon(r) $
and $Z(r)$ are the local mass density and disk half-thickness,
respectively].

The infall speeds of the neutrals ($\vnr$, {\it solid} curves) and
the ions ($\vir$, {\it dashed} curves) are displayed as functions of $r$ for
times $t_{1}$ through $t_{10}$ in Figure 2. (For L1544, the isothermal
sound speed $C=0.21~\rm{km}~{\rm s}^{-1}$.) In the outer regions of the core,
the infall reaches `asymptotic' values for $t_{j} \simgt t_{3}$. We also note that,
for times $t_{j} \simgt t_{6}$ the relative drift speed between the ions
and the neutrals becomes greater in the innermost flux tubes,
reflecting effective ambipolar diffusion (due to depletion of ions
onto grains; see CM93 and CM94 for a discussion) even during the later
stages of the dynamical collapse of the core. This results in
a continued redistribution of the mass and magnetic flux in the
central flux tubes. $M/\Phi_{\rm B}$ increases significantly in the
inner core for $t_{j} \simgt t_{6}$ (the behavior is similar to that
exhibited in Fig. $1c$ of Ciolek \& K\"{o}nigl 1998), and the
mass-to-flux ratio throughout the core is not well-approximated as
being constant at these later times.
Further discussion on the role of ambipolar diffusion during
the later stages of core collapse, such as when a core approaches the
formation of a central point mass (i.e., a protostar) and its
subsequent evolution can be found in Basu (1997) and
Ciolek \& K\"{o}nigl (1998). 

We now apply our results to the L1544 core. Assume that the
symmetry axis of our disk-cloud is inclined at an angle $\theta$ with
respect to the plane of the sky.
A schematic diagram of the assumed geometry is presented in Figure 3.
For an axisymmetric disk with radius
$r$ and half-thickness $Z$, the projected axial ratio (apparent minor
axis/apparent major axis) seen by an observer is
\begin{equation}
\label{axial}
q = \sin \theta + \frac{Z}{r} \cos \theta. 
\end{equation}
(Note that, for
$\theta = 90\arcdeg$, $q = 1$, i.e., a disk viewed
face-on, while, for $\theta = 0\arcdeg$, $q = Z/r$, i.e.,
the disk is seen edge-on.) Maps of the gas and dust distribution in
L1544 (e.g., T98, W99; Ward-Thompson et al. 1999) find $q$ in the range
0.46 to 0.6. We adopt a value of 0.53. From the numerical simulation we
find that, for the region of the core in the range
$0.02~{\rm{pc}} \simlt r \simlt 0.1~\rm{pc}$, $Z(r)/r$ ranges from
0.24 to 0.3. For specificity, we use $Z(r)/r \simeq$ 0.27. From this value
equation (\ref{axial}) implies a tilt angle $\theta \simeq 16\arcdeg$ (or,
equivalently, the cloud is inclined with respect to the line of sight
by an angle $\Psi = 90\arcdeg - \theta = 74 \arcdeg$). From this
geometry it follows that the observed maximum 
line-of-sight infall will be reduced by the factor
$\cos 16\arcdeg = 0.96$. 

In Table 1 we list some of our model predictions for several
physical quantities at two different radii in the L1544 core,
as well as the recent values measured at the same positions by T98 and
W99. The values we list for our numerical results
are taken from the curves labeled $t_{3}$ in Figures $1a$-$1c$;
most quantities change very little for $t_{j} > t_{3}$.
The only notable change at later times is in $|\vnr|$ and $|\vir|$
at $r=0.02$ pc; they increase there by 32\% and 26\% respectively.
As mentioned in \S~2, the maximum resolution of the W99 core survey
($\sim 0.01$ pc), although greater than in previous surveys, would
still be unable to resolve the inner central density peak in our model,
since, for $t_{j} \simeq t_{3}$, the radius of this central region is
$5.6 \times 10^{-3}~\rm{pc}$ (see Fig. $1a$).
Our choice of the data at time $t_3$ as possibly representing the
current evolutionary state of L1544 seems to be reasonably consistent
with recent sub-mm observational maps, which resolve nearly uniform column
densities on scales $\sim 10^3~\rm{AU}$. The line-of-sight number
column density in the inner (`flat') region of our model's core at
$r = 1.2 \times 10^3~\rm{AU}$ is
$N_{\rm{los}} = (\sign/\sin 16\arcdeg)(1/m_{\rm n})= 2.6 \times 10^{23} \,
(\sign/\sin 16 \arcdeg)(2.33 {\rm{a.m.u.}}/m_{\rm n}) = 1.3 \times 10^{23}~\rm{cm}^{-2}$;
for comparison, Andr\'{e} et al. (1999, see their Fig. 2) measure
$N_{\rm{los}} \approx 10^{23}~\rm{cm}^{-2}$ in L1544 at this radial
distance. Moreover, Ward-Thompson et al. (1999, see their Table 2)
find the density in L1544 on these scales to be $\approx 10^6~\cc$;
by contrast, at time $t_3$ our model has $\nn = 2.2 \times 10^6~\cc$
at this position.

In Table 1 we have also included $|\vir|$ at 0.14 pc, and 
$|\vnr|$ at 0.02 pc, neither of which have yet been reported. We note
that the ion-neutral drift speed,
$\vd \equiv \vir - \vnr = 0.03~\rm{km}~{\rm s}^{-1}$ at 0.02 pc and
$0.04~\rm{km}~{\rm s}^{-1}$ at 0.14 pc, is in agreement with our
earlier models (see \S~2), as well as the observed upper limit of drift
speeds in dense cores ($\approx 0.03~\rm{km}~{\rm s}^{-1}$), as deduced
by Benson et al. (1998). Examination of Table 1 shows the model results
to be in excellent agreement with those values observed in L1544.

The appearance of a circularly symmetric function when tilted by our 
predicted $\theta \simeq 16 \arcdeg$ is shown in Figure 4 at time
$t_{3}$. The contours can represent lines of constant density or column
density (similar contours of a model fit to the observed column density
are presented in Fig.  2$b$ of W99), or even constant values of the
infall speed $|v|$ at any given vertical ($z$) layer of the cloud. If
the cloud is infinitesimally thin, the observed line of sight velocity
at any position has magnitude
$|v_{\rm los}| = |v| \cos \theta \sin \varphi$,
where $\varphi = \arctan(|z^\prime/x^\prime|)$, and $x^\prime$ and
$z^\prime$ are sky coordinates as depicted in Figures 3 and 4. This
yields $|v_{\rm los}| = 0$ along the major axis of the cloud. However,
in actuality our model cloud has a finite thickness $Z$, hence any line
of sight through the tilted cloud will pick up some $v_{\rm los}$
if the observed spectral line is optically thin or has its optical
depth unity surface within the cloud. A calculation of the predicted 
spectra in this geometry for various tracers and/or viewing angles 
is more involved, and will be discussed at a later time.

Our model also makes predictions about other physical quantities in
L1544. Figure 5
shows the radial profile of the magnetic field strength in the
equatorial plane of the cloud, $\Beq$ at the same eleven times $t_j$
as in Figure $1a$. At the time of core formation, the central magnetic
field strength $B_{\rm{z,c,crit}} = 19.4~\muG$.
For $t_{j} \simgt t_3$, $\Beq$ is equal to $71~\muG$ at $r=0.02~\rm{pc}$, and
$21~\muG$ at $0.14~\rm{pc}$. The line-of-sight magnetic field strengths
are reduced by the factor $\sin 16\arcdeg= 0.28$, hence Zeeman
measurements would measure a field strength $=5.79~\muG$ at 0.14pc, and
$19.6~\muG$ at 0.02 pc. By simultaneously and self-consistently solving
the rate equations for the abundances of charged species, (see CM93 
and CM95 for discussions), we calculate the total degree of ionization
$\xxi$ (= atomic ion + molecular ion abundances) as a function of
position in the cloud; in this model, $\xxi=1.1 \times 10^{-8}$ at
0.02 pc, and $\xxi = 5.6 \times 10^{-8}$ at 0.14 pc.

Finally, our calculation allows us to infer the lifetime of the L1544
core. The evolutionary timescale $\tevol$ for this model (i.e., the
time taken to go from $t_{0}$ to $t_{10}$, essentially the time to form
a central protostar) is $\simeq 2.7~\Myr$. Since the central
mass-to-flux ratio becomes critical at a time $\tcrit = 1.3~\Myr$, the
actual amount of time that our L1544 model exists as a {\em
supercritical} core is $\tlife = \tevol - \tcrit = 1.4~\Myr$. The 
{\em detectable} lifetime for ammonia observations is
$\tdetect = \tevol - \tammonia$, where $\tammonia$ is
the time (corresponding to $n_{\rm n,c} \geq 1.1 \times 10^4~\cc$) at 
which ammonia becomes collisionally excited in the core. For our model,
$\tammonia = 1.5~\Myr$, yielding $\tdetect = 1.2~\Myr \, (\approx \tlife$). 
This is compatible with the $\sim 1~\Myr$ lifetime of starless cores
advocated in the recent study of Lee \& Myers (1999).
\vspace{-4ex}
\section{Summary}
\vspace{-2ex}
Recent studies have presented detailed, high-resolution
observations of the L1544 core. These observations play an important
role in constraining and testing theoretical models of star formation
in molecular clouds. In light of these new observations, we have
reexamined previously published ambipolar diffusion models of the
formation and collapse of supercritical cores in magnetically supported
molecular clouds. We find that, despite the questions raised in the
papers associated with these surveys, the general characteristics of
supercritical cores in the earlier ambipolar diffusion models are
consistent with physical features exhibited in the L1544 core. We have
also presented a new numerical simulation designed to more accurately
reproduce observed features of L1544, as well as make predictions for
heretofore unobserved quantities. The model uses reasonable input
parameters well within the physically allowed range. Comparing our
results to those measurements obtained in the recent studies, we find
the resulting mass and velocity distribution of our theoretical
model to be in excellent agreement with values deduced at two distinct
radial distances, 0.02 pc and 0.1 pc (e.g., see Table 1). There is also
agreement with the mass and column densities deduced by sub-mm
measurements of dust emission on smaller scales. This model allows us to
deduce the orientation of L1544 with respect to the plane of the sky
($\theta \approx 20\arcdeg$), and makes predictions about the
line-of-sight magnetic field strength
($B \sin 20\arcdeg \approx 20~\muG$) and degree of ionization
($\xxi > 10^{-8}$) at different points in the core. The detectable
lifetime of the collapsing core in our L1544 model ($\simeq 1.2~\Myr$)
is comparable to that inferred from statistics of an ensemble of cores
(Lee \& Myers 1999).

While our model supercritical core is to be compared with the infall zone of 
the L1544 core, of radial extent $\sim 0.1$ pc, our cloud model also yields   
numbers for the subcritical envelope. However, we have not presented these as
predictions for the envelope structure. As discussed in our earlier
papers, the outer density structure is strongly dependent on adopted
initial conditions. For example, a sharp break in the density profile, with
power-law index $s$ dropping significantly lower than $-2$ just outside the
supercritical core, is a feature of models with highly subcritical
envelopes (e.g., model 7 of BM95b, and model $\rm{B_{UV}}$ of CM95). Further
modeling of the envelope structure can be guided by new data for the outer regions 
(e.g., see preliminary data in Andr\'{e} et al. 1999).

Hopefully, our new study will stimulate greater interaction between
observation and theory, in which observations can more tightly
constrain input parameters for theoretical models, and the models can
make new predictions for unobserved quantities. Although the occurrence
of processes not included in our model cannot be ruled out, our results
suggest that current observations of infall in starless cores are
compatible with the evolution of cores with supercritical mass-to-flux
ratio that have formed through ambipolar diffusion.

\acknowledgements{GC gratefully acknowledges support
from the NY Origins of Life Center (NSCORT) and Physics Dept. at RPI,
under NASA grant NAG5-7598. SB's work was supported by the Natural
Sciences and Engineering Research Council of Canada. We would also like to
thank Philippe Andr\'{e}, Phil Myers, and Derek Ward-Thompson, as
well as an anonymous referee, for helpful comments and discussions.}

\newpage
\begin{table}
\begin{center}
TABLE 1 \\
{\sc Physical Quantities in the L1544 Core \\}
\begin{tabular}{lll}
\hline
\hline
\mbox{\hspace{6em}} & Predicted${}^{\rm a}$ & Observed \\
\hline
$r \simeq$ 0.14 pc : ${}^{\rm b}$ &    & \\
\hspace{1em} $M$ &  $7.7~\msol$ & $8~\msol$ \\
\hspace{1em} $|\vnr|$ & $0.11~\rm{km}~\rm{s}^{-1}$ & $0.1~\rm{km}~{\rm s}^{-1}$ \\
\hspace{1em} $|\vir|$ & $0.07~\rm{km}~{\rm s}^{-1}$ & ------ \\
\hspace{1em} $s = \frac{\partial \ln \nn}{\partial \ln r} $ & $-1.7$ & $\sim -1.5$ \\ 
\hspace{1em} $B_{\rm los}$ & $5.8~\muG$ & ------ \\ \\
$r \simeq$ 0.02 pc : ${}^{\rm c}$ & & \\
\hspace{1em} $M$   & $0.90~\msol$ & $1.2~\msol$ \\
\hspace{1em} $|\vnr|$ & $0.14~\rm{km}~\rm{s}^{-1}$ & ------ \\
\hspace{1em} $|\vir|$ & $0.11~\rm{km}~{\rm s}^{-1}$ & $0.08~\rm{km}~{\rm s}^{-1}$ \\
\hspace{1em} $\nn$ & $3.3 \times 10^5~\cc$ & $4 \times 10^5~\cc$ \\
\hspace{1em} $B_{\rm los}$ &  $19.6~\muG$ & ------ \\ \\
\hline
\end{tabular}
\end{center}
\mbox{\hspace{8em}}${}^{\rm a}$ Assuming a disk tilt-angle
$\theta = 16\arcdeg$. \\
\mbox{\hspace{8em}}${}^{\rm b}$ Observed values taken from T98. \\
\mbox{\hspace{8em}}${}^{\rm c}$ Observed values taken from W99.
\end{table}
\newpage
\begin{center}{\bf Captions to Figures} \end{center}
\figcaption{Spatial profiles of physical quantities, as functions
of radius $r$ at eleven different times
$t_{j}$ ($j$ = 0, 1, 2,...,10), chosen such that the central
density at time $t_{j}$ is a factor $10^j$ greater than the
initial central density. These times are, respectively, 0, 2.27, 2.60,
2.66, 2.680, 2.684, 2.685, 2.6856, 2.68574, 2.68577, and
2.68578 $\Myr$. An asterisk on a curve, present only after the
formation of a supercritical core ($t > t_{0}$), locates the
instantaneous radius of the critical magnetic flux tube.
An open circle on a curve marks the instantaneous position of
the central thermal critical ($\simeq$ Jeans) lengthscale.
({\it a}) Density.
({\it b}) The exponent $s \equiv \partial \ln \nn/\partial \ln r$.
({\it c}) Vertical column density.
}
\figcaption{Radial component of neutral velocity ($\vnr$, {\it solid}
curves) and ion velocity ($\vir$, {\it dashed} curves).
Note that the isothermal sound speed $C$ for this model is $0.21~\rm{km}~{\rm s}^{-1}$. 
The times $t_{j}$ and the labels on the curves are the same as
in Figs. $1a$-$1c$.}
\figcaption{Schematic diagram showing the assumed geometry and
orientation of a disklike model cloud, with radius $R$ and
half-thickness $Z$. The disk symmetry axis, which is aligned with
the $z$-axis, is slanted at an angle $\theta$ with respect to the
$z^\prime$-axis, which is taken to lie in the plane of the
sky.}
\figcaption{Appearance of physical quantities in the L1544
core at time $t_{3}$, projected onto axes in the plane of the sky. The curves depict
the location of loci of constant physical quantities, such as column
density, density, and infall speed (directed toward the
origin), for our assumed inclined model cloud.
For the column density, the contours represent, respectively
(starting from the outermost contour), $2.6 \times 10^{21}~\rm{cm}^{-2}$,
$3.6 \times 10^{21}~\rm{cm}^{-2}$, $4.8 \times 10^{21}~\rm{cm}^{-2}$,
$6.7 \times 10^{21}~\rm{cm}^{-2}$, and
$1.2 \times 10^{22}~\rm{cm}^{-2}.$
For the density, the contours correspond to (in the same order)
$1.2 \times 10^{4}~\cc$, $2.3 \times 10^{4}~\cc$, $4.0 \times 10^{4}~\cc$,
$8.1 \times 10^{4}~\cc$, and $2.7 \times 10^{5}~\cc$; for the
infall speed they would represent contours with
$-0.10~\rm{km}~\rm{s}^{-1}$, $-0.12~\rm{km}~\rm{s}^{-1}$,
$-0.127~\rm{km}~\rm{s}^{-1}$, $-0.136~\rm{km}~\rm{s}^{-1}$, and
$-0.140~\rm{km}~\rm{s}^{-1}$.
Similar contours are presented for a model fit column density 
in Fig. $2b$ of W99.}
\figcaption{Profile of the vertical component of the magnetic field
in the equatorial plane of the cloud. All times and labels are
the same as in the preceding figures.}

\setcounter{figure}{0}
\newpage
\begin{figure}
\plotone{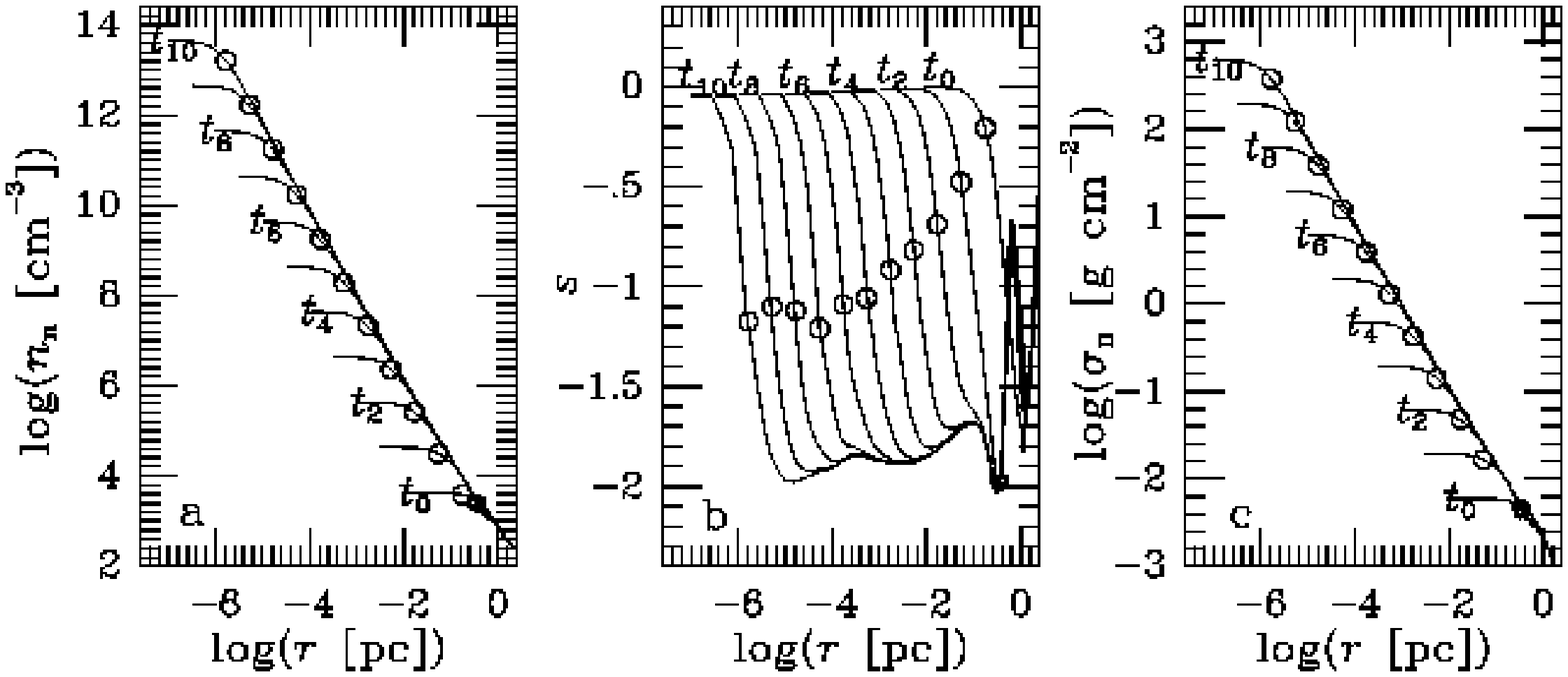}
\caption{}
\end{figure}
\begin{figure}
\plotone{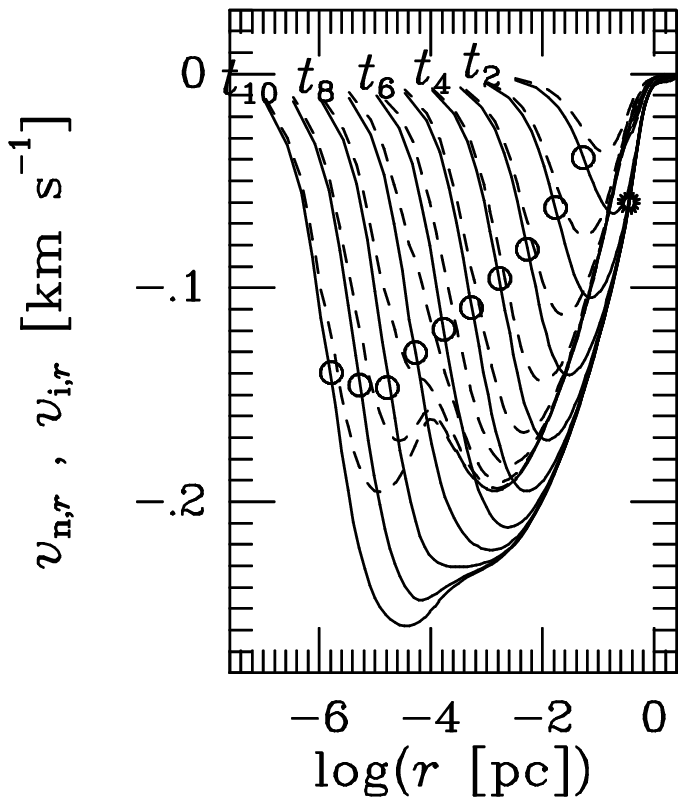}
\caption{}
\end{figure}
\begin{figure}
\plotone{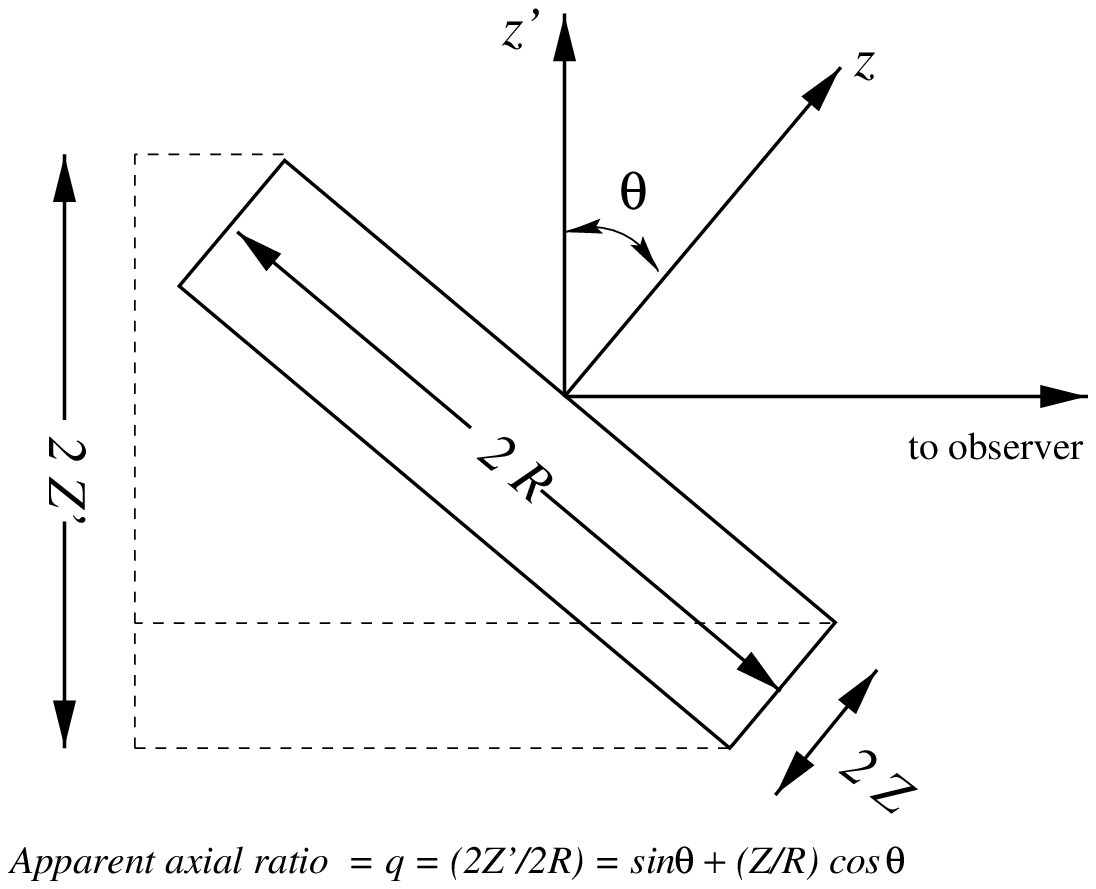}
\caption{}
\end{figure}
\begin{figure}
\plotone{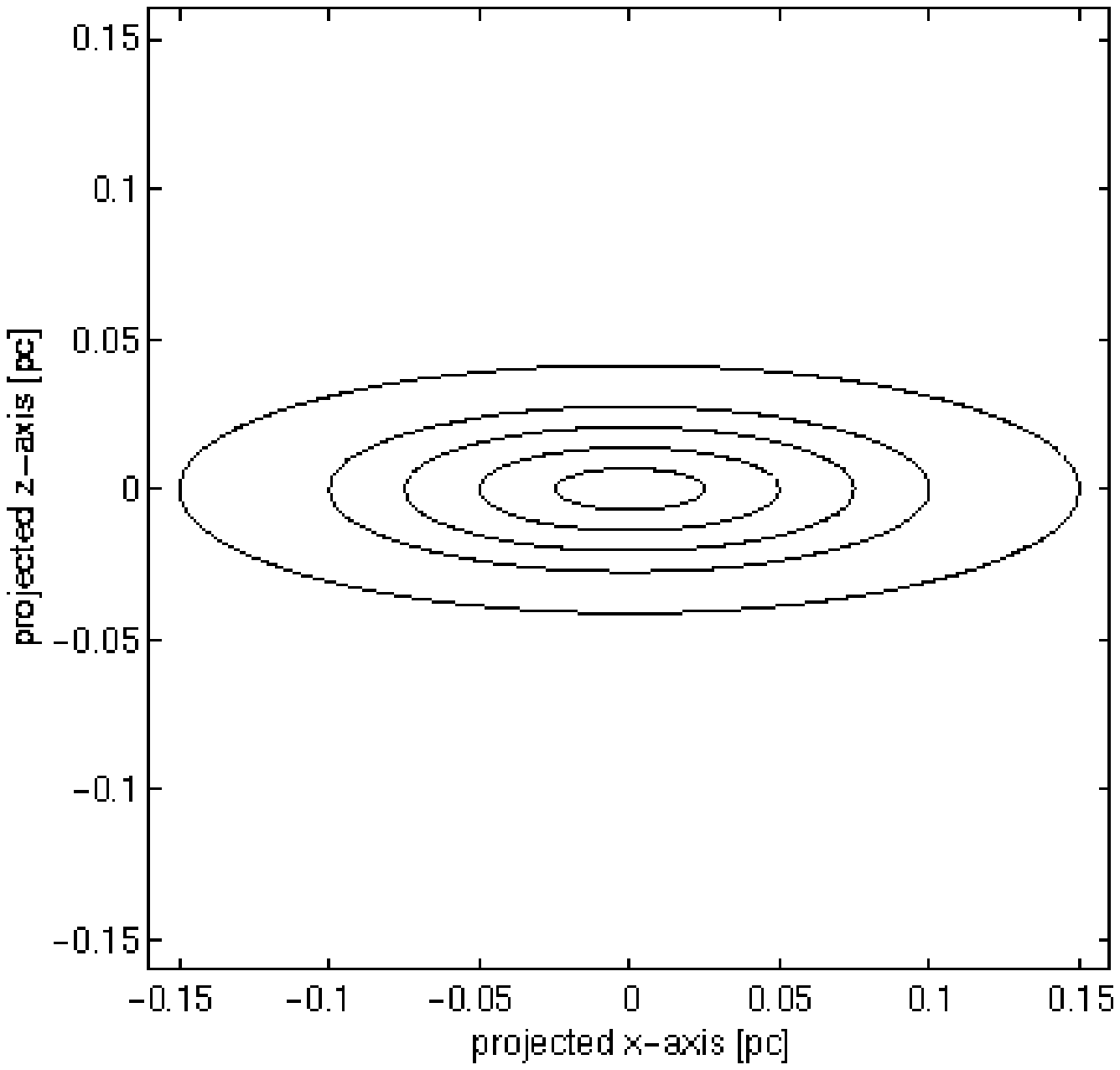}
\caption{}
\end{figure}
\begin{figure}
\plotone{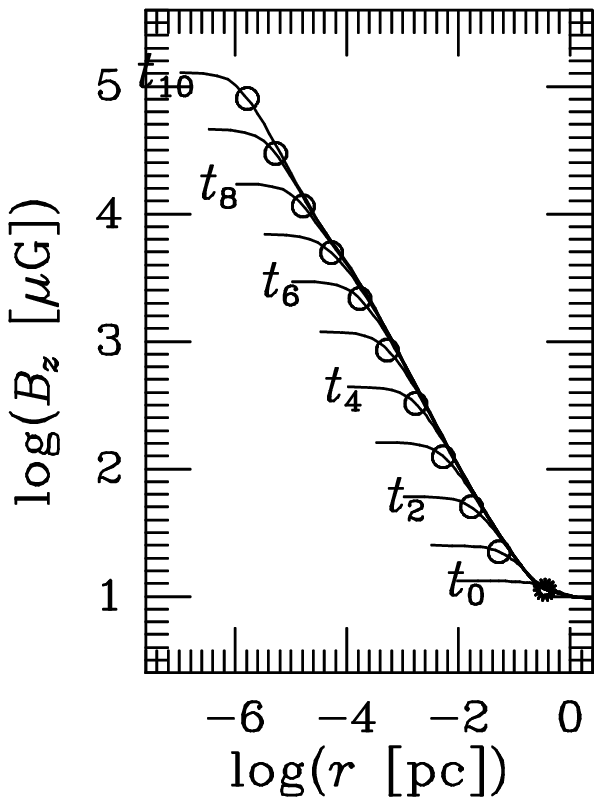}
\caption{}
\end{figure}

\end{document}